\documentstyle[epsf]{article}

\title{
Numerical analyses of the nonequilibrium electron transport
through the Kondo impurity beside the Toulouse point
}
\author{Yoshihiro Nishiyama \\
{\it Department of Physics, Faculty of Science,
Okayama University} \\
{\it Okayama 700-8530, Japan}
}
\date{(Received \hspace*{50mm})}

\begin{document}

\maketitle

E-mail: nisiyama@psun.phys.okayama-u.ac.jp

TEL: +81-86-251-7809

FAX: +81-86-251-7830

\section*{Abstract}

Nonequilibrium
electron transport through the Kondo impurity
is investigated numerically for the
system 
with twenty conduction-electron levels.
The electron current under finite voltage drop
is calculated in terms of the 
`conductance viewed as transmission' picture proposed by Landauer.
Here, we take into account the full transmission processes of both
the many-body correlation and the hybridization amplitude
up to infinite order.
Our results demonstrate, for instance,
how the exact solution of the differential conductance
by Schiller and Hershfield
obtained at the Toulouse point
becomes deformed by more realistic interactions.
The differential-conductance-peak height is suppressed below $e^2/h$
with the width hardly changed through reducing the
Kondo coupling from the Toulouse point,
whereas it
is kept unchanged by further increase of the coupling.
We calculated the nonequilibrium local Green function as well.
This clarifies the spectral property of the Kondo impurity
driven far from equilibrium.

{\bf PACS codes/keywords}: 72.15.Qm (Scattering mechanisms and Kondo effect),  
82.20.Mj (Nonequilibrium kinetics),                                      
75.40.Mg (Numerical simulation studies)

\section{Introduction}
\label{section_Introduction}

Recently, in fine semiconductor devices,
the Kondo effect has been observed
at very low temperatures \cite{Goldhaber-Gordon98,Cronenwett98,Schmid98}.
The indications of the Kondo effect are
drawn mainly from the electron-transport measurements
under various controlled system parameters
such as the
temperature, the gate voltage, the hybridization amplitude, 
the magnetic field and
the bias voltage.
In particular, the differential conductance
${\rm d}I(V)/{\rm d}V$ measured over a wide range of the bias voltage $V$
shows a distinctive indication of the Kondo effect:
The differential conductance shows a peak
around the zero bias ($V=0$)
with the width comparable
to the Kondo temperature.
The peak height grows as the temperature is decreased.
These features had been observed in various other
experimental realizations
such as tunnel metallic junctions
 and metallic point contacts \cite{Wyatt64,Logan64,Gregory92,Ralph94},
although
in these systems, 
the system parameters are not so freely
tunable as in the semiconductor devises.
Nevertheless,
the series of experiments has been stimulating
theoretical investigations so far
\cite{Appelbaum66,Anderson66,Glazman88,Ng88,Kawabata91,Kawabata98,%
Meir91,Meir92,Meir93,Ng93,Wingreen94,Hershfield91,Hershfield92,Schiller95,%
Majumdar98,Schiller98,Inoshita93,Izumida98,Oguri95,Oguri97,Konig96}.
The essence of the differential-conductance feature
is attributed mainly to the shape of
the local density of states at the impurity.
Note that the density of states
contributes to the transmission probability in the intermediate stage of
the transmission processes.
It is well known that the density of states at the Kondo impurity
is of a peak structure with the width comparable
to the Kondo temperature.
This picture yields comprehensive qualitative understanding of
experimental observations.

Apart from the practical interest such as to
explain the essential features of experimental observations,
the subject may cast a rather fundamental
problem
where the many-body correlation and 
the nonequilibrium-driving force coexist, and are both important.
Without the biquadratic correlation, 
we can calculate the nonequilibrium transport exactly \cite{Caroli71}
with the Keldysh approach \cite{Keldysh65}.
On the other hand, the transport
coefficient for correlated system
is given by the Kubo formula, although the
application is restricted in the vicinity of
equilibrium.
There have not yet been found any frameworks to
command the nonequilibrium transport and the
biquadratic correlation simultaneously. 
In fact,
in the above mentioned theories,
a number of perturbative approaches ---
either the voltage drop or the coupling to the leads are
supposed to be infinitesimal ---
are employed.
Recently, Schiller and Hershfield, however,
succeeded in treating the situation rigorously \cite{Schiller95,Schiller98}.
To the best of our knowledge,
their solution is the first exact solution of a nonequilibrium
correlated system.
They found that the 
nonequilibrium Kondo model becomes integrable 
at the so-called Toulouse point \cite{Toulouse70},
where the transverse and longitudinal
Kondo couplings to both leads are tuned very carefully
so as to
satisfy the Emery-Kivelson condition \cite{Emery92}.
Their result at zero temperature shows that the differential conductance
is of a pure Lorentzian shape.
This result might be rather disappointing, because the
pure Lorentzian form had been known to be
realized in simple quadratic system \cite{Caroli71}.
This rather uncharacteristic feature may be due to the
fact that the system at the integrable point is
nothing but quadratic.
In fact, in a subsequent paper \cite{Majumdar98},
they considered some perturbations
to the Toulouse point, and tried to show how the integrable-point
result is deformed by these perturbations.

Here, we formulate the nonequilibrium transport phenomenon
with many-body correlation
in a manner that is suitable for numerical simulation.
Based on the formulation,
we
investigate the transport through the Kondo impurity.
The transport coefficient
is evaluated in terms of the `conductance viewed as the
transmission' picture proposed by Landauer \cite{Landauer57,Imry99}.
Here, we take into account the full transmission processes
of both the many-body correlation and
the hybridization amplitude up to infinite order.
The rest of this paper is organized as follows.
In the next section, we explain the model Hamiltonian,
the so-called resonance level model, which is equivalent to
the Kondo model \cite{Schlottmann82,Guinea85}.
The equivalence is explained, and the advantages to investigate the former
are explicated.
In Section \ref{section_formulation}, the formulation of the nonequilibrium 
transport phenomenon, mentioned above briefly, is presented in detail.
We also propose the algorithm to calculate the nonequilibrium Green 
function.
In Section \ref{section_results}, we give the results of numerical simulation.
In the last section, we give summary of this paper.

\section{Resonance level model}
\label{section_model}

In this section, we explain the so-called resonance level model.
We simulated this model rather than the Kondo model.
Although both models are equivalent \cite{Schlottmann82,Guinea85},
it is much more advantageous
to treat the former model rather than the latter.
The Hamiltonian of the resonance level model is given by,
\begin{equation}
\label{Hamiltonian}
{\cal H}=
\sum_k
   \left(v_{\rm F}k+\frac{eV}{2}\right)L^\dagger_k L_k
+
\sum_k
   \left(v_{\rm F}k-\frac{eV}{2}\right)R^\dagger_k R_k
+
V_{\rm h}(S^+ c+S^- c^\dagger)
+
U S^z\frac{c^\dagger c - c c^\dagger}{2}.
\end{equation}
The operator $L^\dagger_k$ ($R^\dagger_k$) creates a
conduction electron of wave number $k$ in the left (right) lead.
The wave number $k$ is distributed uniformly over the range
$k=-k_{\rm c} \sim k_{\rm c}$
($k_{\rm c}=\pi$); here,
the wave number is indexed in such a way as
$v_{\rm F} k_i=2i/(N-1)-2/(N-1)-1$ ($i=1,2,\cdots,N$).
That is, there are $N$ conduction-electron levels in each lead.
The parameter $v_{\rm F}k_{\rm c}(=\omega_{\rm c})$ 
is regarded as the unit of energy throughout this paper;
namely, $v_{\rm  F}k_{\rm c}=1$.
The operator $c^\dagger$ creates an electron at the impurity position;
that is,
\begin{equation}
c^\dagger=\frac{1}{\sqrt{2N}}\sum_k (L^\dagger_k+R^\dagger_k).
\end{equation}
The operators $S^z$ and $S^\pm$ are the conventional $S=1/2$
spin operators.
Hence, the parameters $V_{\rm h}$ and $U$ denote the hybridization
amplitude and the density-density correlation strength
to the conduction-electron bands, respectively.
Because we study the nonequilibrium situation
where both leads are kept half-filled,
the parameter $eV$ give the chemical potential difference between
two leads.
The way how we realize such nonequilibrium situation
in computer
will be shown in the next section.

% In the case of equilibrium $eV=0$,
It is shown that the above model (\ref{Hamiltonian}) is equivalent
to the anisotropic Kondo Model
by means of the bosonization technique \cite{Schlottmann82,Guinea85}.
The anisotropic Kondo model is given by,
\begin{eqnarray}
\label{Kondo_model}
{\cal H}_{\rm K} &=&
\sum_{k\sigma}
   \left(v_{\rm F}k+\frac{eV}{2}\right)L^\dagger_{k\sigma} L_{k\sigma}
+
\sum_{k\sigma}
   \left(v_{\rm F}k-\frac{eV}{2}\right)R^\dagger_{k\sigma} R_{k\sigma}
                    \nonumber   \\
& & +
J_\parallel S^z \frac{c^\dagger_\uparrow c_\uparrow 
      -c^\dagger_\downarrow c_\downarrow}{2}
+
J_\perp (S^+ c^\dagger_\downarrow c_\uparrow 
  +S^- c^\dagger_\uparrow c_\downarrow),
\end{eqnarray}
where the index $\sigma$ denotes the spin index; $\sigma=\uparrow$
or $\downarrow$.
According to the bosonization analysis, the following
mapping relations should hold \cite{Guinea85};
\begin{equation}
\label{mapping_relation}
V = \frac{\rho_k J_\perp \omega_{\rm c}}{2}(\rho_k \omega_{\rm c})^{-1/2} 
             ,
\end{equation}
and,
\begin{equation}
U = (1-\sqrt{2}|1-\rho_k J_\parallel/2|) /2\rho_k,
\end{equation}
with the density of states $\rho_k=(2\pi v_{\rm F})^{-1}$.
The mapping relations tell the followings.
First, at
$J_\parallel=2(1-1/\sqrt{2})/\rho_k$,
there is a very special point at which the parameter $U$ vanishes.
(Note, however, that quantitative reliability of these mapping relations
is {\it skeptical}.
Precise connection between these models
is somewhat lost in the bosonization procedure.
Therefore, for instance, we cannot tell definitely at which point ($V_{\rm h}$-$U$)
the isotropic Kondo point is realized.)
At this point, the resonance revel model (\ref{Hamiltonian}) becomes
quadratic, and this point, the so-called Toulouse point \cite{Toulouse70},
corresponds to the strong-coupling fixed point of the Kondo model.
Second, the reduction of the Kondo coupling from the Toulouse point
corresponds to the attractive biquadratic correlation $U<0$,
whereas the furthermore increase of the Kondo coupling is viewed
as the repulsive correlation $U>0$.
Roughly speaking, the previous rigorous analysis is
concerned in the Toulouse point \cite{Schiller95,Schiller98}.
In the analysis, however, the authors used the model with {\it two channel}
density-density-correlation
term, and arrived at the Toulouse point of the {\it two channel} Kondo model.
Because our simulation is rather incapable of the two-channel Kondo physics,
we have chosen the model with the conventional Kondo coupling.
We investigate numerically the parameter region beside the
Toulouse point $U=0$ in Section \ref{section_results}.

We summarize below the advantages to treat the resonance level model
rather than the Kondo model:
First, in the former model, 
the spin index is
dropped so that we can diagonalize the system size twice as large
as that of the Kondo model. 
The physical reason of this dropping is originated in the followings.
Because the Kondo problem is essentially of a one-dimensional problem,
The charge and spin degrees of freedom are separated completely.
Therefore, only the spin degrees of freedom are subjected
to the impurity-spin scattering.
That is, the spin degrees of freedom
 transmit through the impurity through the assistance of the scattering,
whereas
the charge degrees of freedom do not transmit at all.
That is why we can ignore the charge sector.
(Therefore, the perfect-transmission conductance is {\it not} $2 e^2/h$, which
arises in conventional models, but $e^2/h$, because the charge-sector transmission
is ignored completely here.)
Secondly,
as is discussed in the previous paragraph,
it is very advantageous that
we can start from the Kondo fixed point, namely, the Toulouse
point, which readily contains the very essence of the Kondo physics.
Hence, we can study the effect of the biquadratic term 
in a systematic manner
as the gradual deviation from the fixed point.

\section{
Formulation of the nonequilibrium electron transport
}
\label{section_formulation}

In this section, we discuss how we set up the formalism
to simulate the nonequilibrium electron-transport phenomenon.
The formalism is prepared in such a way that
it is readily implemented in a computer-simulation algorithm.
First, we give a conductance formula based on the
`conductance viewed as transmission'
picture
proposed by Landauer \cite{Landauer57,Imry99}.
Extensive use of numerical technique
enables the evaluation of
the transition-matrix elements containing
higher order contributions of both the hybridization amplitude and 
the density-density correlation.
Then, we give the algorithm to calculate the nonequilibrium
Green function, which is useful in order to investigate the
spectral property under nonequilibrium current flow.

According to the Landauer picture,
the electron current is given by the transmission probability
between two leads;
\begin{eqnarray}
I&=&e\sum_{kk'}
\left\{
f(v_{\rm F}k)
\frac{2\pi}{\hbar}
|T_{kk'}|^2(1-f(v_{\rm F}k'))\delta(v_{\rm F}k-v_{\rm F}k'+eV)
\right. \nonumber \\
& &
\left.
-
(1-f(v_{\rm F}k))
\frac{2\pi}{\hbar}
|T_{kk'}|^2 f(v_{\rm F}k')\delta(v_{\rm F}k-v_{\rm F}k'+eV)
\right\}   \nonumber \\
&=& e N^2 \rho_\varepsilon^2 \int {\rm d}\varepsilon
\left\{
f(\varepsilon-eV)
\frac{2\pi}{\hbar}
|T_{\varepsilon-eV , \varepsilon}|^2\left(1-f(\varepsilon)\right)
-
\left(1-f(\varepsilon-eV)\right)
\frac{2\pi}{\hbar}
|T_{\varepsilon-eV ,\varepsilon}|^2f(\varepsilon)
\right\}       \nonumber \\
\label{current_formula}
&=& e N^2 \rho_\varepsilon^2 \int_{-eV/2}^{eV/2} {\rm d}\varepsilon
\frac{2\pi}{\hbar}
|T_{\varepsilon-eV/2 , \varepsilon+eV/2}|^2               .
\end{eqnarray}
We have used
the Fermi-distribution function $f(\varepsilon)$ and
the density of states $\rho_\varepsilon=1/2$, and supposed
that the system is at 
zero temperature.
$T_{k k'}$ denotes the transition-matrix element
from the wave number $k$ of the left lead to $k'$ of the right lead.
The explicit expression of $T_{kk'}$ is considered afterwards.
Thereby, we obtain the differential conductance,
\begin{eqnarray}
G(eV)&=&\frac{{\rm d}}{{\rm d}V} I    \nonumber \\
&=& \frac{{\rm d}}{{\rm d}V}
 e N^2 \rho_\varepsilon^2 \int_{-eV/2}^{eV/2} {\rm d}\varepsilon
\frac{2\pi}{\hbar}
|T_{\varepsilon-eV/2 , \varepsilon+eV/2}|^2             .
\end{eqnarray}

Therefore,
the conductance $G(eV)$ is given by
the derivative of the integral of 
the following integrand,
\begin{equation}
g(\varepsilon) =
 e N^2 \rho_\varepsilon^2
\frac{2\pi}{\hbar}
|T_{\varepsilon-eV/2,\varepsilon+eV/2}|^2            .
\end{equation}
(Note that the function $g(\varepsilon)$ is
an even function because of the particle-hole symmetry.) 
Hence, it is expected that the integrand $g(\varepsilon)$
yields the conductance,
\begin{equation}
\label{differential_conductance}
G(eV) \simeq e g(eV/2) ,
\end{equation}
unless the integrand depends on $eV$ very much.
Fortunately, it is known that in the noninteracting case $U=0$,
the current is expressed in terms of the following rigorous form \cite{Caroli71},
\begin{equation}
\label{Caroli_formula}
I=\frac{2\pi e (V_{\rm h}/\sqrt{2})^4}{\hbar}
  \int_{-eV/2}^{eV/2} {\rm d}\omega |{\cal G}(\omega)|^2 \rho_\varepsilon^2  .
\end{equation}
(That is, at $U=0$, the transition matrix 
$T_{\varepsilon-eV/2,\varepsilon+eV/2}$ is related to the local Green
function at the impurity site ($\propto {\cal G}(\omega=\varepsilon)$).)
Because the integrand is 
{\em completely} independent on $eV$,
the relation (\ref{differential_conductance}) holds exactly.
We expect that beside $U=0$, the relation continues to hold well.
In fact, as is presented afterwards in Section 4.3,
$\omega(=\varepsilon)$-dependence is dominated and
determines the essential feature (peak structure)
of the integrand,
whereas $eV$ just causes sub-dominant detailed contributions.
As is shown below, a handy formula (\ref{WML_formula})
for conductance estimate exists and is
expressed in the similar form as ours (\ref{current_formula}).
In the presence of $U$,
the integrand behaviors 
of ours (\ref{current_formula}) and the formula (\ref{current_formula})
become
{\em qualitatively}
different.
Therefore, at the present stage,
it would be well worth considering the conductance
in the approximation level of (\ref{differential_conductance}).

% Suppose that the integrand hardly depends on the voltage drop $eV$.
% We arrive at the compact formula,
% \begin{equation}
% \label{differential_conductance}
% G(eV) \approx
%  e^2 N^2 \rho_\varepsilon^2
% \frac{2\pi}{\hbar}
% |T_{0,eV}|^2                                  .
% \end{equation}
% This assumption is actually satisfied at the quadratic point $U=0$
% \cite{Caroli71}.
% Nevertheless, the readers who are not satisfied by the assumption
% may regard the formula (\ref{differential_conductance})
% as the integrand of the
% current formula (\ref{current_formula}).
% In our scheme, of course, we could have differentiated 
% numerically the
% electron current.
% Yet, it is more advantageous to consider the integrand
% itself
% (as the differential conductance), 
% because we need to compare the above first-principle
% result with approximative one, which is also given by an integral formula;
% see below.
% Hence, in order to make the comparison clearer,
% we focused our attention on the integrand of the electron-current formula
% (\ref{current_formula}),
% and expect that this would also yield a very precise estimate of the
% differential conductance.

Next, we calculate the transition-matrix element.
In order to do that, we must prepare the initial and
final states,
\begin{equation}
|{\rm i}\rangle=L^\dagger_{k_{N/2+1}} |g_0\rangle,
\end{equation}
and
\begin{equation}
|{\rm f}\rangle=R^\dagger_{k_j} |g_0\rangle
\ \ (j=N/2+1,N/2+2,\cdots,N),
\end{equation}
respectively.
We have supposed that $N$ is an even integer.
The state $|g_0\rangle$ represent the situation
where both leads are half-filled;
\begin{equation}
|g_0\rangle = L^\dagger_{k_1} L^\dagger_{k_2} \cdots L^\dagger_{k_{N/2}}
  R^\dagger_{k_1} R^\dagger_{k_2} \cdots R^\dagger_{k_{N/2}}
   | 0 \rangle.
\end{equation}
In the single-particle transmission process, the energy should be conserved.
This conservation condition restricts 
the available values of the voltage drop
within
the series
$eV= 2(j-N/2-1)/(N-1)$ ($j=N/2+1,N/2+2,\cdots,N$);
note that
our conduction-electron band is discrete.
With use of these initial and final states,
we obtain the
transition probability,
\begin{equation}
\label{golden_rule}
w_{{\rm i} \rightarrow {\rm f}}=
\left.
\frac{{\rm d}}{{\rm d}t} |c_{\rm f} (t)|^2
\right|_{t=0},
\end{equation}
with,
\begin{equation}
\label{jikan_hatten}
c_{\rm f} (t)=
\langle {\rm f} |
\frac{
   U_{{\rm I}\eta}(t,-\infty) |{\rm i}\rangle
     }
     { \left| U_{{\rm I}\eta}(t,-\infty) |{\rm i}\rangle \right| } .
\end{equation}
In the above, $U_{{\rm I}\eta}$ is the time-evolution operator of the 
interaction representation;
\begin{equation}
U_{{\rm I}\eta}(t_2,t_1)={\rm T}{\rm e}^{-{\rm i}\int_{t_1}^{t_2}
{\rm d}t {\cal H}_1(t) /\hbar }   ,
\end{equation}
with,
\begin{equation}
{\cal H}_1(t)={\rm e}^{{\rm i} {\cal H}_0 t/\hbar}
\left(
V_{\rm h}(S^+ c+S^- c^\dagger)
+
U S^z\frac{c^\dagger c - c^\dagger c}{2}
\right)
{\rm e}^{-{\rm i}({\cal H}_0 + {\rm i}\eta) t/\hbar} ,
\end{equation}
and,
\begin{equation}
{\cal H}_0=
\sum_k
   \left(v_{\rm F}k+\frac{eV}{2}\right)L^\dagger_k L_k
+
\sum_k
   \left(v_{\rm F}k-\frac{eV}{2}\right)R^\dagger_k R_k .
\end{equation}
(${\cal H}={\cal H}_0 + {\cal H}_1$.)
The normalization factor in eq. (\ref{jikan_hatten})
is vital, because the infinitesimal damping parameter $\eta$ violates
the unitarity of $U_{{\rm I}\eta}$.
This fact is known as the Gell-Mann-Low theorem \cite{Gell-Mann51}.
Moreover, it is notable
that $\eta$ plays a significant role to realize
`nonequilibrium dissipative state' breaking the time-reversal
symmetry \cite{Hershfield93}.
Using the property of the time-evolution operator $U_{{\rm I}\eta}$,
we obtain the following expression,
\begin{eqnarray}
w_{{\rm i}\rightarrow{\rm f}}  & = &  \frac{{\rm d}}{{\rm d}t}
    \left.
       \frac{
  |\langle {\rm f} | U_{{\rm I}\eta}(t,-\infty) | {\rm i} \rangle|^2
            }
            {
  \langle{\rm i}|U_{{\rm I}\eta}(\infty,t)
             U_{{\rm I}\eta}(t,-\infty)|{\rm i}\rangle
            }
    \right|_{t=0}                 \nonumber \\
& = &
 \frac{{\rm d}}{{\rm d}t}
    \left.
       \frac{
  |\langle {\rm f} | U_{{\rm I}\eta}(t,-\infty) | {\rm i} \rangle|^2
            }
            {
  \langle{\rm i}|U_{{\rm I}\eta}(\infty,-\infty))|{\rm i}\rangle
            }
    \right|_{t=0}                 \nonumber \\
& = &
\frac{1}{|U_{{\rm I}\eta}(0,-\infty)|{\rm i}\rangle|^2}
\left.
 \frac{{\rm d}}{{\rm d}t}
 |\langle {\rm f} | U_{{\rm I}\eta}(t,-\infty) | {\rm i} \rangle|^2
\right|_{t=0}    .
\end{eqnarray}
Through expanding the operator $U_{{\rm I}\eta}$ into the Dyson series,
one arrives at the following formula,
\begin{equation}
w_{{\rm i} \rightarrow {\rm f}}=\frac{2\pi}{\hbar}
\left|\langle {\rm f} | T | {\rm i} \rangle\right|^2
   \delta(E_{\rm f}-E_{\rm i})   ,
\end{equation}
where the transition matrix $T$ is,
\begin{equation}
\label{transition_matrix}
T=\frac{1}{\left|U_{{\rm I}\eta}(0,-\infty) |{\rm i}\rangle \right|}
{\cal H}_1
\left(
1-\frac{1}{{\cal H}_0+{\cal H}_1-E_{\rm i}-{\rm i}\eta} {\cal H}_1
\right)     ,
\end{equation}
with ${\cal H}_0|{\rm i}\rangle=E_{\rm i}|{\rm i}\rangle$ and,
\begin{equation}
\label{time_evolution}
U_{{\rm I}\eta}(0,-\infty)|{\rm i}\rangle=
\left(
1-\frac{1}{{\cal H}_0+{\cal H}_1-E_{i}-{\rm i}\eta} {\cal H}_1   
\right) |{\rm i}\rangle          .
\end{equation}
The above formulae complete our technique to compute the
electron conductance; note the relation 
$T_{0,eV}=\langle{\rm f}|T|{\rm i}\rangle$.
Readers may have noticed that we are subjected to the Lippmann-Schwinger
formulation essentially \cite{Hershfield93}.
It is notable that the formula for the transition matrix 
(\ref{transition_matrix})
contains {\em infinite-order contributions} with respect to
${\cal H}_1$,
and thus our simulation does {\em not} resort to any perturbative
treatments.

Lastly, we show the way to calculate the non-equilibrium
Green function,
\begin{eqnarray}
\label{Green_function}
{\cal G}(eV,\omega)&=&\int {\rm d}t {\rm e}^{{\rm i}\omega t} {\cal G}(eV,t)
                  \\
{\cal G}(eV,t)&=&
  -{\rm i}\Theta(t)\langle g_{eV} | 4\{ S^-(t) , S^+ \} | g_{eV} \rangle
                      ,
      \nonumber
\end{eqnarray}
where the nonequilibrium steady state is given by,
\begin{equation}
| g_{eV} \rangle = \frac{U_{{\rm I}\eta}(0,-\infty)|g_0\rangle}
 {|U_{{\rm I}\eta}(0,-\infty)|g_0\rangle|}  .
\end{equation}
Because of the same reasoning as the above, we need a normalization factor.
The time-evolved state is calculated similarly,
\begin{equation}
U_{{\rm I}\eta}(0,-\infty)|g_0\rangle = 
 \left(
   1-\frac{1}{{\cal H}_0+{\cal H}_1-E_{g0}-{\rm i}\eta}{\cal H}_1
 \right)
   |g_{0}\rangle .
\end{equation}
This gives an explicit expression for the state $|g_{eV}\rangle$.
What is still left is to calculate the Fourier transform of the
time correlation function of eq. (\ref{Green_function}).
At first glance, one might think that it is impossible,
because
all the eigenstates are needed to evolve the time correlation.
Gagliano and Balseiro, however, invented a way to express
the Green function (\ref{Green_function}) in 
a compact continued-fraction
form with the Lanczos tri-diagonal elements \cite{Gagliano87}.
Through utilizing their technique, the scheme to calculate
nonequilibrium Green function is completed.
We emphasize that in our scheme, we do not resort to
any perturbative treatments.

The Green function is useful to estimate approximate value
of electron current.
% So far, two formulae are known for estimating the current;
So far, the following formula has been used to obtain 
the nonequilibrium current;
\begin{equation}
\label{WML_formula}
I=\frac{e}{\hbar}
 \int_{-eV/2}^{eV/2} {\rm d}\omega 
  \frac{\Gamma^L \Gamma^R}{\Gamma^L + \Gamma^R}
   \cdot
   \frac{-1}{\pi}{\rm Im} {\cal G}(eV,\omega) ,
\end{equation}
with the Abrikosov-Suhl resonance frequency for each lead \cite{Meir92,Meir93},
namely,
\begin{equation}
\label{AS_frequency}
\Gamma^{{\rm L},{\rm R}}=2\pi\rho_{\varepsilon} (V_{\rm h}/\sqrt{2})^2   .
\end{equation}
As is mentioned in Introduction, this formula (\ref{WML_formula}) insists
that the spectral density at the impurity site ($\propto {\rm Im} {\cal G}$)
contributes to the current flow.
Note that
as in our above-mentioned formula (\ref{current_formula}), the current is
expressed in terms of an integral form.
An so,
because of the same reasoning mentioned before,
the integrand of (\ref{WML_formula})
does hardly depend on $eV$ for $U \approx 0$, yielding a compact expression
for the differential conductance;
\begin{equation}
\label{differential_conductance_WML}
G(eV) \simeq \frac{e^2}{\hbar} \cdot
  \frac{\Gamma^L \Gamma^R}{\Gamma^L + \Gamma^R}
   \cdot
   \frac{-1}{\pi}{\rm Im} {\cal G}(eV,\omega=eV/2)    ,
\end{equation}
In the next section, we report that the conductance estimate
with our formula (\ref{differential_conductance}) differs
qualitatively from that of the above handy relation.

\section{Numerical results and discussions}
\label{section_results}
So far, we have prepared prescriptions
for simulating the nonequilibrium electron transport numerically.
In this section, we carry out numerical simulations
based on the prescriptions.

\subsection{Preliminaries of our numerical simulation}
Here, we summarize details of our numerical computation.
We simulated the system consisting of twenty conduction-electron
levels.
(The form of the Hamiltonian is explained in 
Section \ref{section_model}; see eq. (\ref{Hamiltonian}).)
That is, there are ten levels for each lead; $N=10$.
(Owing to the presence of the density-density-correlation
term in the Hamiltonian
(\ref{Hamiltonian}), the Hamiltonian-matrix elements
are not so sparse as in conventional
diagonalization simulations.
That costs considerable computation time, even though the
dimensionality of the Hilbert space ($\sim 2^{21}$)
is not so particularly overwhelming.)

The most important computation stage is that to evaluate the
resolvent which is appearing
in eqs. (\ref{transition_matrix}) and (\ref{time_evolution}). 
The resolvent is computed with the conjugate-gradient algorithm.
The convergence of the conjugate-gradient iteration is not very stable.
This instability becomes more serious as the coupling to the
leads is strengthened.
This instability has been encountered so far in general in calculating
resolvent of many-body problem numerically.
Here, we have used the following new trick to overcome this difficulty.
We substituted the parameter $\eta$ in
eqs. (\ref{transition_matrix}) and (\ref{time_evolution})
with,
\begin{equation}
\label{eta_formula}
\eta= \eta^{(0)} + \eta^{(2)} \sum_k ((v_{\rm F}k)^2 L^\dagger_k L_k
                 +(v_{\rm F}k)^2 R^\dagger_k R_k)   .
\end{equation}
Now,
$\eta$ is not a c-number constant, but is an operator.
Because the parameter determines the energy resolution,
it should be of the order of the conduction-band
discrete spacing.
Hence, we set $\eta^{(0)}=0.1$.
The first term alone results in a desperate instability 
of numerical procedure.
Hence, we have added
the second new term.
The meaning of this term may be transparent:
The term enforces the life time of each 
conduction-band level to vary as 
the square of the
excitation energy measured from the Fermi level.
Such dependence might be reasonable, because
the Landau theory for the interacting electron gas
concludes the same dependence of the life time.
Here, we set $\eta^{(2)}=0.06$ for $V_{\rm h}=0.3$.

We found that this choice of parameters is the best one:
Reader may feel that
our $N=10$-level approximation 
of the conduction band might be serious, because the Fermi-%
surface singularity is important to realize the Kondo effect.
Yet, owing to the resonance-level-model mapping,
we are actually in the {\em strong}-Kondo-coupling regime (Toulouse point),
where the Kondo temperature is given by
$\sim \Gamma^{\rm L}+\Gamma^{\rm R}=2\pi\rho_{\varepsilon}V_{\rm h}^2$;
see eq. (\ref{AS_frequency}).
As a matter of fact, our choice of 
$V_{\rm h}=0.3$ gives the Kondo temperature $\sim 0.28$
which is larger than
the conduction-band level spacing $2/N=0.2$.
On the contrary, the energy-resolution broadening term $\eta$ is not
negligible compared with the Kondo temperature so that
we are suffered from a broadening of the zero-bias-anomaly peak.
We found that for $V_{\rm h}>0.3$, the numerical instability becomes
desperately serious, and exceedingly large smearing factor $\eta$ is needed.
Hence, we concentrated ourselves
on the above-mentioned parameter point with $V_{\rm h}=0.3$,
$\eta^{(0)}=0.1$ and $\eta^{(2)}=0.06$, where we can achieve the
Kondo effect and barely manage the numerical instability 
with a modest energy-resolution broadening.

% For strong-coupling region $V_{\rm h}>0.3$, 
% the above-mentioned numerical instability arises very severely,
% whereas for weak-coupling region $V_{\rm h}<0.3$, the Abrikosov-Suhl resonance frequency 
% becomes less than our conduction-bond spacing. 
% % and
% %$\eta^{(2)}=0.08$ for $V_{\rm h}=0.4$.
% %Those values are the minimal for each $V_{\rm h}$ within the
% %bounds of stable convergence.
% %Nevertheless,
% %it is expected that the choice of $\eta$ might
% %not so significant,
% %because it only concerns the energy resolution.

\subsection{Differential conductance}
In this subsection, we present our numerical result
of the differential conductance.
The calculation is based on the formula (\ref{differential_conductance}).
We discuss the result with an emphasis on the effect of $U$,
namely, the deviation from the Toulouse point.

\begin{figure}[htbp]
\begin{center}\leavevmode
\epsfxsize=17cm
\epsfbox{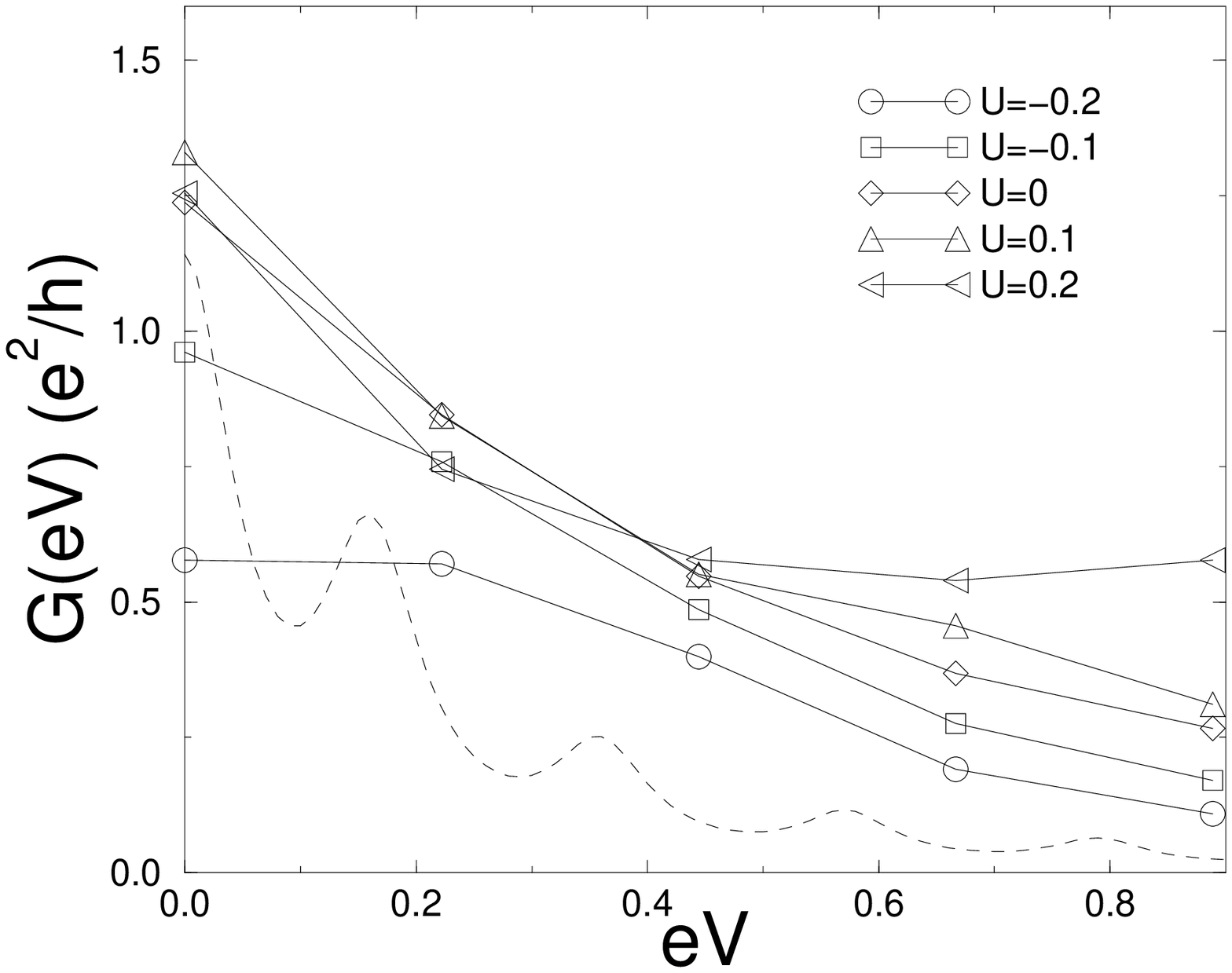}
\end{center}
\caption{
Differential conductance ({\protect \ref{differential_conductance}})
is plotted for $V_{\rm h}=0.3$ and various $U$.
The dashed curve shows the result with the analytic formula
%({\protect \ref{differential_conductance_Caroli}}) that is valid at $U=0$.
({\protect \ref{differential_conductance_WML}}) that is valid at $U=0$.
}
\label{ogon_Vt3}
\end{figure}

In Fig. \ref{ogon_Vt3}, we plotted the differential conductance
against the voltage drop $eV$
for $V_{\rm h}=0.3$ and various $U$.
We see that the differential conductance is maximal
at the zero bias $eV=0$
irrespective of $U$.
This peak structure
is known as the
`zero-bias anomaly.' 
The anomaly has been observed experimentally, and
is very significant to confirm
that the electron
actually 
tunnels between two leads with a certain quantum-mechanical
tunneling amplitude.

First, we discuss the data for $U=0$
(the Toulouse point).
At this quadratic case $U=0$, an analytical prediction for the
differential conductance is available;
see Section \ref{section_formulation}.
According to that,
the differential conductance is
given by the formula
(\ref{differential_conductance_WML}) with
the Green function,
\begin{equation}
\label{free_Green_function}
{\cal G} (\omega)=\frac{1}{
\omega+\delta - 
   \sum_k\frac{(V_{\rm h}/\sqrt{2N})^2}{\omega+\delta-v_{\rm F}k-eV/2}
              -
   \sum_k\frac{(V_{\rm h}/\sqrt{2N})^2}{\omega+\delta-v_{\rm F}k+eV/2}
                  }    .
\end{equation}
The result of this formula with $\delta=0.05$
is also plotted
as a dashed line in Fig. \ref{ogon_Vt3}.
% (Only the curve of $eV=0$ is plotted, because at
% $U=0$, the Green function becomes almost independent on $eV$.)
We see that our numerical result succeeds in reproducing the
zero-bias conductance $\sim e^2/h$, which is accordant with the Landauer
theory, and the zero-bias peak as well.
On the contrary, our zero-bias-peak width is broadened 
substantially compared with
the analytical prediction curve.
That is due to the energy-resolution-smearing factor $\eta$
(\ref{eta_formula}),
which cannot be omitted in order to stabilize the 
conjugate-gradient procedure.
(This factor $\eta$ broadens the energy resolution, and loosens
the energy-conservation constraint.
Thereby,
this parameter enhances the
transmission probability, resulting in an
enhanced estimate of conductance.)
As is explicated in Section 4.1,
we had implemented a new trick that
the factor $\eta$
(life time) is not a constant, but
is gained near the band edges.
Therefore, owing to $\eta$, our conductance data
suffer a correction and are enhanced 
especially for large potential drop $eV \sim 1$.
Hence, we rather focus our attention on the qualitative variation of 
the conductance curve $G(eV)$ 
due to the introduction of the many-body correlation $U$.
It should be stressed that,
as is explained in Section \ref{section_formulation},
besides this broadening, our simulation is free from any approximations, and
actually, takes into account full many-body correlation processes.

% Because of the discreteness of our conduction-band spectrum,
% the analytical formula gives a wavy differential-conductance
% curve, and yields rather
% suppressed value of conductance.
% We see that our numerical result succeeds in reproducing
% several important aspects of the prediction:
% Our result yields the
% zero-bias
% conductance $\sim e^2/h$, which is accordant with the Landauer theory.
% The differential conductance decreases as the bias voltage
% is increased.
% In the whole range of $eV$, our simulation result
% yields rather larger value of conductance than 
% that of the analytical formula.
% This enhancement
% might be attributed to the parameter $\eta$ given by eq. (\ref{eta_formula})
% appearing in the resolvents
% (\ref{transition_matrix}) and (\ref{time_evolution}).
% This factor broadens the energy resolution, and loosens
% the energy-conservation constraint.
% Thereby,
% This parameter enhances the
% transmission probability, resulting in an
% enhanced estimate of conductance.
% Because of the same mechanism,
% our conductance is enhanced to a similar extent also for $U \ne 0$.

Second, keeping that in mind,
let us turn to the case of $U<0$.
Note that the case $U<0$ corresponds to the situation where the Kondo
coupling is reduced from the value of
the Toulouse point; refer to 
the mapping relations (\ref{mapping_relation}).
In Fig. \ref{ogon_Vt3},
we see that the zero-bias conductance is suppressed 
significantly by the reduction of the
Kondo coupling.
It is very surprising, because 
conventional picture
based on the one-particle description,
namely, eqs. (\ref{differential_conductance_WML}) and
(\ref{AS_frequency}), gives the conductance
$e^2/h$ irrespective of the hybridization amplitude as far as the
left-right hybridization couplings are symmetric.
Therefore, we notice that
such reduction is of a nontrivial many-body effect, and in principle,
that should be managed 
by {\it ab initio} treatments such as ours.
As a matter of fact, a conductance estimate based on the
convenient formula (\ref{differential_conductance_WML}) gives
qualitatively opposite behavior
that the conductance is {\em enhanced} with $U<0$;
this discrepancy is reported in the next subsection.

% One might have expected, at first glance,
% that the zero-bias conductance would not be
% affected by $U(<0)$.
% %In fact, the Kondo impurity form the so-called local Fermi liquid,
% %where the physics is connected adiabatically with a one-body picture.
% Yet,
% the present simulation indicates that the one-body description
% is no longer valid, and
% many-body correlation causes a very significant contribution to the
% transport coefficient.
Furthermore, it seems that
the width of the zero-bias-anomaly peak is not changed very much by $U<0$.
It may be interesting to compare our findings with
those of
the lowest-order-%
perturbation theory with respect to the Toulouse point
by Majumdar, Schiller and Hershfield \cite{Majumdar98}.
They studied the effect of certain three types of perturbations
to the differential conductance.
These perturbations are thought to drive the integrable Toulouse Hamiltonian
in the direction of the weak Kondo coupling;
namely, more realistic coupling. 
(Because they are employing a field-theoretical description
together with a very sophisticated canonical transformation,
the origin of these
perturbations in the language of the original Kondo Hamiltonian
is rather unclear, and is to be clarified.)
They found that some two perturbations keep the height fixed, but
sharpen the width, whereas the other perturbation suppresses the height
with the width unchanged.
The true effect is speculated to be a certain mixture of those effects.
% We clarified here that the reduction of the Kondo coupling
% reduces the height, but does not change
% the width very much.
Our results suggest that the reduction of the Kondo coupling
reduces the height, but does not change
the width very much; that is, in the field-theoretical 
language \cite{Majumdar98}, the latter type of perturbation
might be realized actually in the physics of $U<0$.
% In order to address more definite conclusion for this issue,
% however, more accurate simulation might be desirable;
% one needs to resolve the instability in the conjugate-gradient procedure.

Next, we discuss the case $U>0$.
This repulsive density-density correlation is
interpreted as the {\em very} strong Kondo coupling, which is 
even exceeding
the Toulouse-coupling strength.
Hence, this case might not be so relevant
in the language of the Kondo physics, and has not yet been 
studied very well \cite{Bedurftig99}.
Nevertheless, apart from the context of the Kondo problem,
this situation is interesting by its own right.
In Fig. \ref{ogon_Vt3},
We observe that
as the repulsive correlation is increased,
no particular change is observed for the differential conductance
especially for $eV \sim 0$. 
That is, despite of the repulsion,
inherent quantum-mechanical resonance restores the
hybridization between the impurity and the conduction electron.
This situation will be more clarified in the next subsection
through referring to the spectral-function data.

% Finally, we show the differential conductance
% for larger value of the hybridization
% amplitude $V_{\rm h}=0.4$ in Fig. \ref{ogon_Vt4}.
% Features are essentially the same as those of 
% the above $V_{\rm h}=0.3$.
% That is,
% the curves exhibit the zero-bias anomaly as well,
% and show rapid suppression of conductance by $U(<0)$.
% it should be noted, however, that
% the zero-bias peak is broadened substantially in this case $V_{\rm h}=0.4$.
% This broadening is consistent with the description of
% eq. (\ref{differential_conductance_WML}),
% because
% the local density of states ($\propto {\rm Im}G(\omega)$)
% is broadened by the hybridization.
% According to the description, the width should increase rapidly as the
% square of the hybridization amplitude.
% As is explained above, the parameter $\eta$ assists the
% tunneling, giving rise to a supplementary enhancement of conductance
% for overall range of the bias voltage.
% This correction seems to grow for
% larger value of $V_{\rm h}$.

\subsection{Nonequilibrium Green function}
In this subsection, we present the numerical result of the
nonequilibrium Green function.
The calculation is based on the formula (\ref{Green_function}).
We concentrate on the imaginary part of the
Green function, which yields the local density of states
at the impurity site;
\begin{equation}
\label{DOS}
\rho(\omega)=-\frac{1}{\pi} {\rm Im} {\cal G} (eV,\omega+{\rm i}\delta)  ,
\end{equation}
with $\delta=0.1$.
Once the spectral function is obtained, one can adopt the
handy formula (\ref{differential_conductance_WML}) to estimate
the conductance.
We show that the estimates (behaviors)
differ qualitatively from our 
first-principle results shown in Section 4.2.

\begin{figure}[htbp]
\begin{center}\leavevmode
\epsfxsize=17cm
\epsfbox{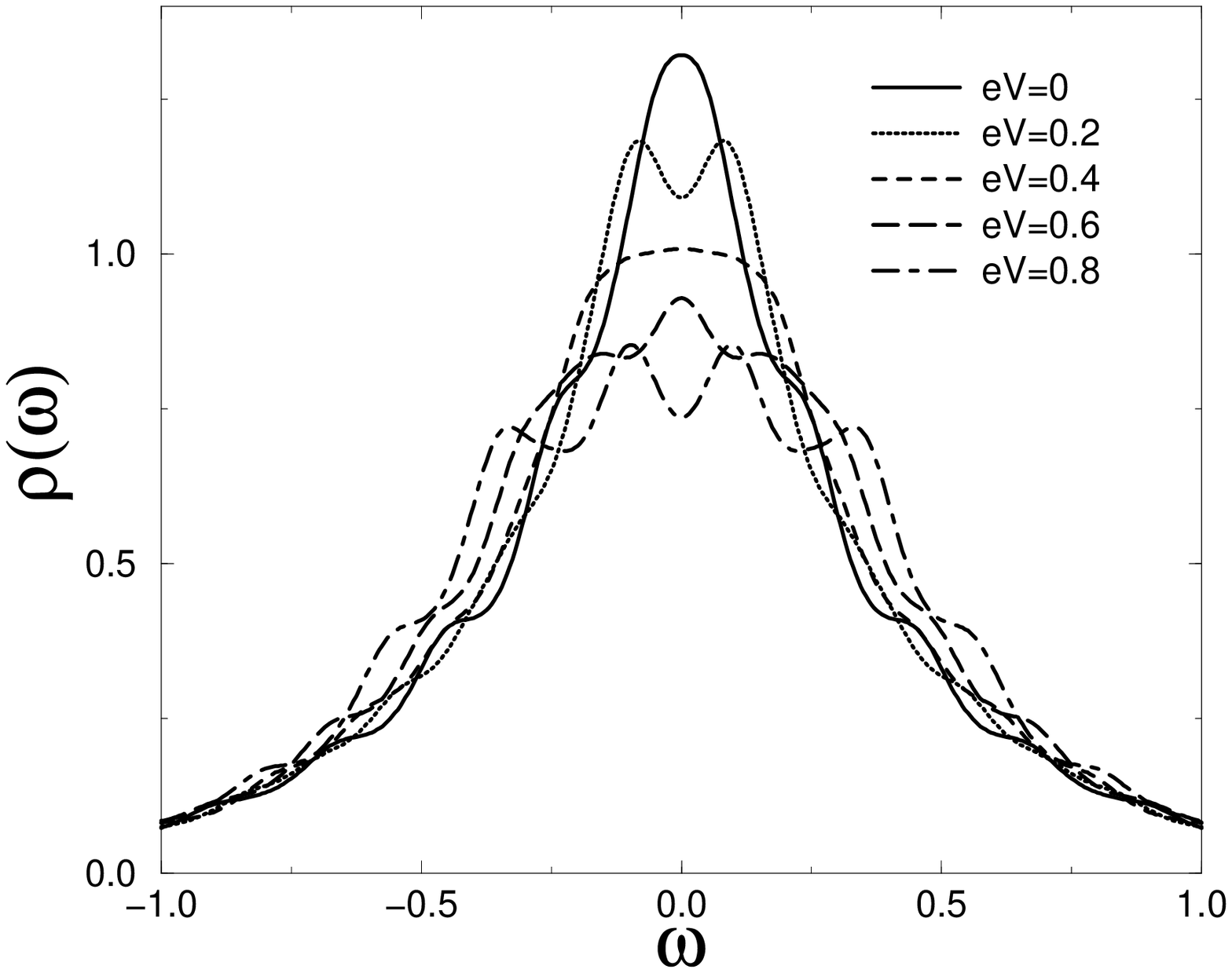}
\end{center}
\caption{
Local density of states at the impurity site
({\protect \ref{DOS}}) for
$V_{\rm h}=0.3$ and $U=0$. 
}
\label{rho_Vt3_Ut001_10}
\end{figure}

\begin{figure}[htbp]
\begin{center}\leavevmode
\epsfxsize=17cm
\epsfbox{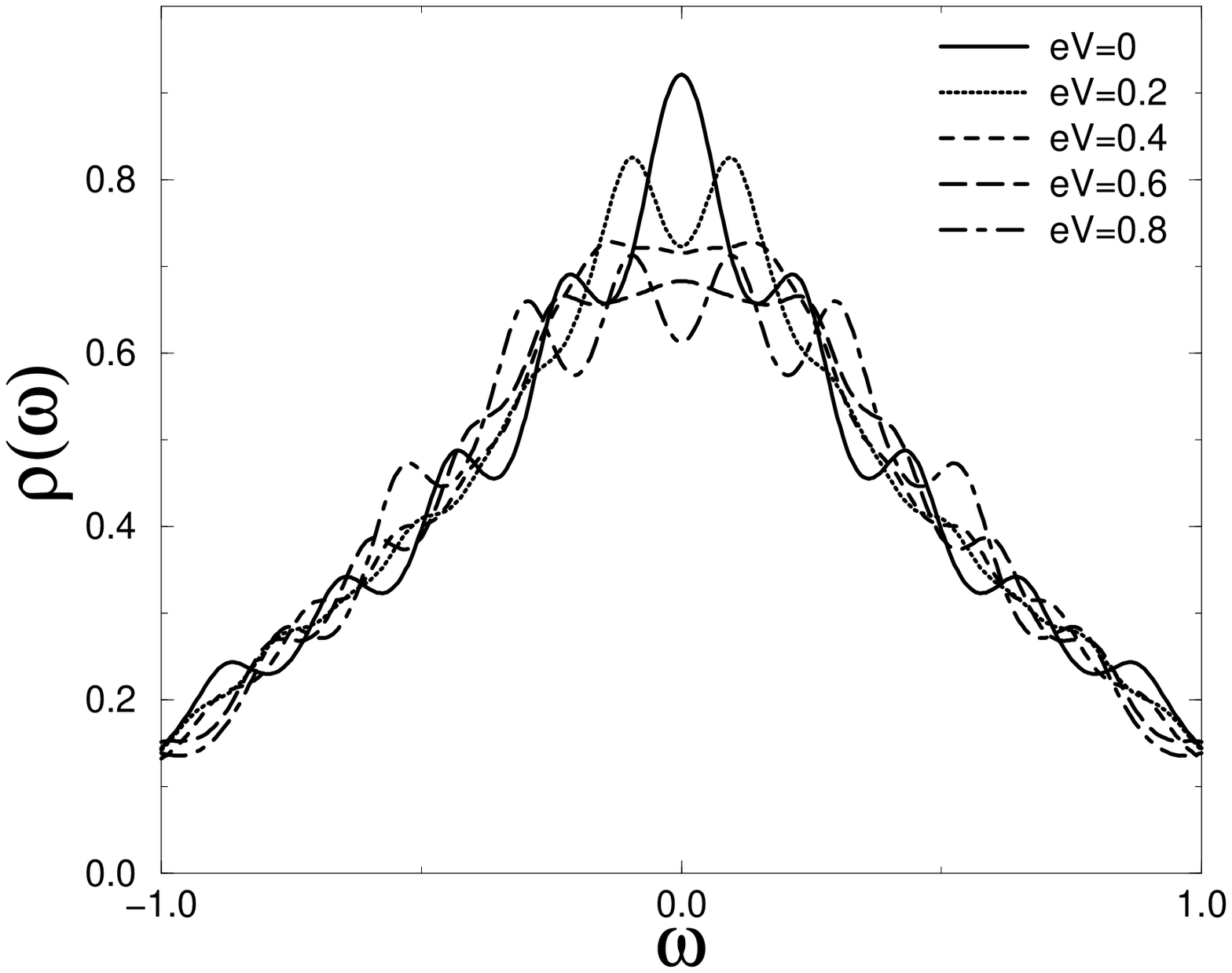}
\end{center}
\caption{
Local density of states at the impurity site
({\protect \ref{DOS}}) for 
$V_{\rm h}=0.3$ and $U=0.2$. 
}
\label{rho_Vt3_Ut2_10}
\end{figure}

\begin{figure}[htbp]
\begin{center}\leavevmode
\epsfxsize=17cm
\epsfbox{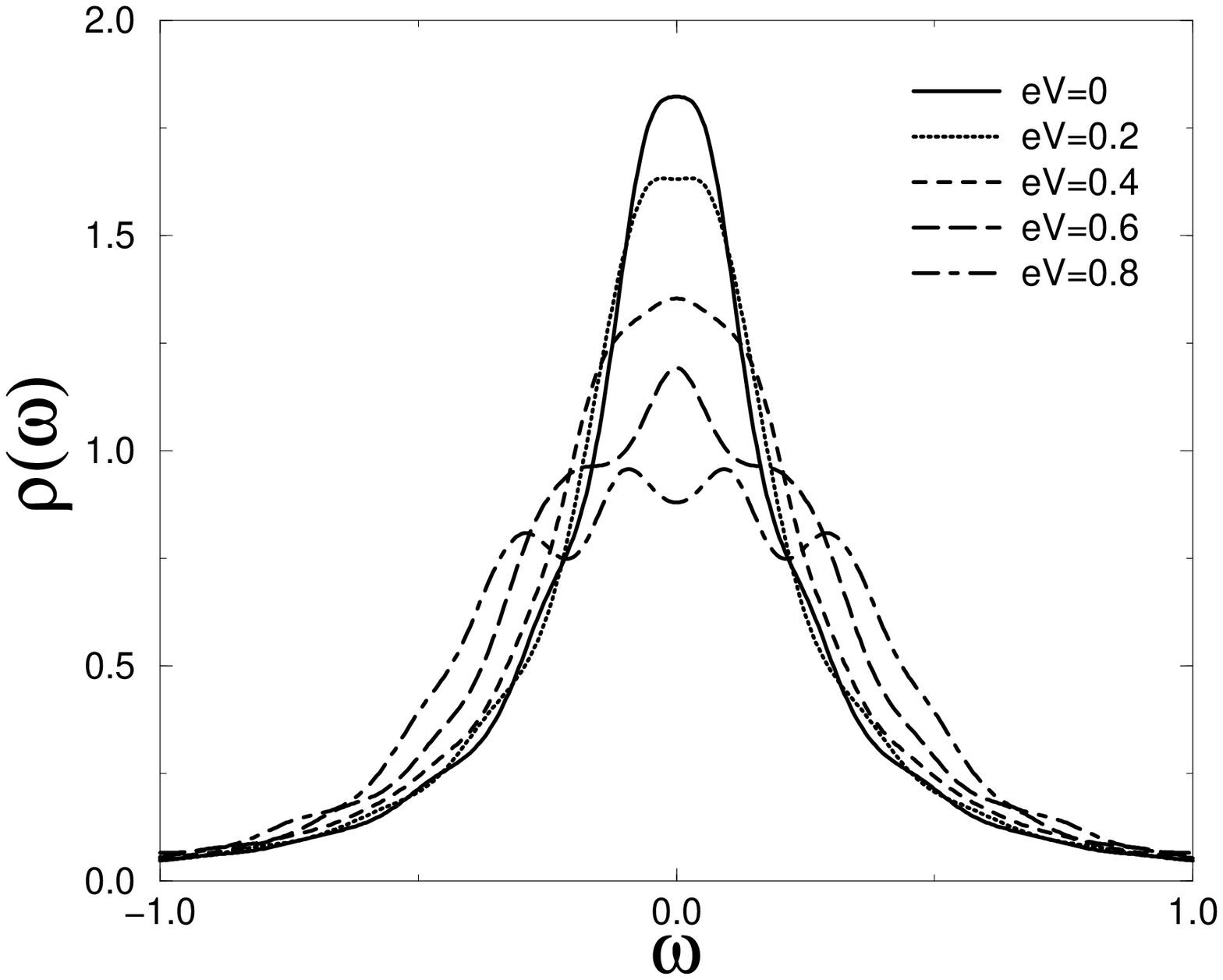}
\end{center}
\caption{
Local density of states at the impurity site
({\protect \ref{DOS}}) for 
$V_{\rm h}=0.3$ and $U=-0.2$. 
}
\label{rho_Vt3_Umt2_10}
\end{figure}

In Figs. \ref{rho_Vt3_Ut001_10}-\ref{rho_Vt3_Umt2_10},
we plotted the density of states
for $V_{\rm h}=0.3$, and $U=0$, $0.2$ and $-0.2$, respectively.
The local density of states shows a peak structure.
That is, the impurity level 
is smeared out by the mixing (resonance)
to the conduction-electron band.
The peak width corresponds to the inverse of the life time of the
impurity state.

First, we discuss the case of $U=0$ (Fig. \ref{rho_Vt3_Ut001_10}).
We see that the local density of states, namely, the resonance
peak,
is suppressed gradually as the bias voltage $eV$ is increased.
That is understood as follows:
For $eV \sim \omega_{\rm c}(=1)$,
the Fermi-energy position approaches the band edge.
In that case,
the resonance is not fully formed because of the disturbance of the 
band edge.
That causes the reduction of the resonance peak for $eV\sim1$.

Second, let us turn to the cases with the many-body
correlation $U \ne 0$.
In those cases, the Green function becomes far more nontrivial,
and contains significant informations.
For $U=0.2$ (Fig. \ref{rho_Vt3_Ut2_10}),
the density of states forms broader resonance than that of $U=0$.
In particular, the sub-peaks at $\omega=\pm eV/2$ are prominent;
note that
the frequencies $\omega=\pm eV/2$ are the Fermi levels
of the leads.
The sub-peak height stays aloft even for $eV>0$.
Thereby, the sub-peaks 
constitute shoulders of the main peak, resulting in considerable
broadening of the main peak.
This feature has been captured by previous studies
\cite{Meir93,Ng93,Konig96}, and is speculated
to be the very essence
of the nonequilibrium Kondo physics.
We emphasize that the present 
simulation does actually capture this characteristic without resorting
to any approximations.
This prominent resonance may be the precursor of the
unexpected large conductance, 
which we reported in the previous subsection.
On the other hand,
for $U=-0.2$ (Fig. \ref{rho_Vt3_Umt2_10}),
the density of states forms a very narrow peak,
and the shoulders at $\omega= \pm eV/2$ are not grown very much.
That is, in this weak Kondo coupling case, the
Kondo resonance is suppressed to a considerable extent.
This is consistent with the observation of the previous
subsection that the (zero-bias) conductance is suppressed
significantly by $U<0$.

\begin{figure}[htbp]
\begin{center}\leavevmode
\epsfxsize=17cm
\epsfbox{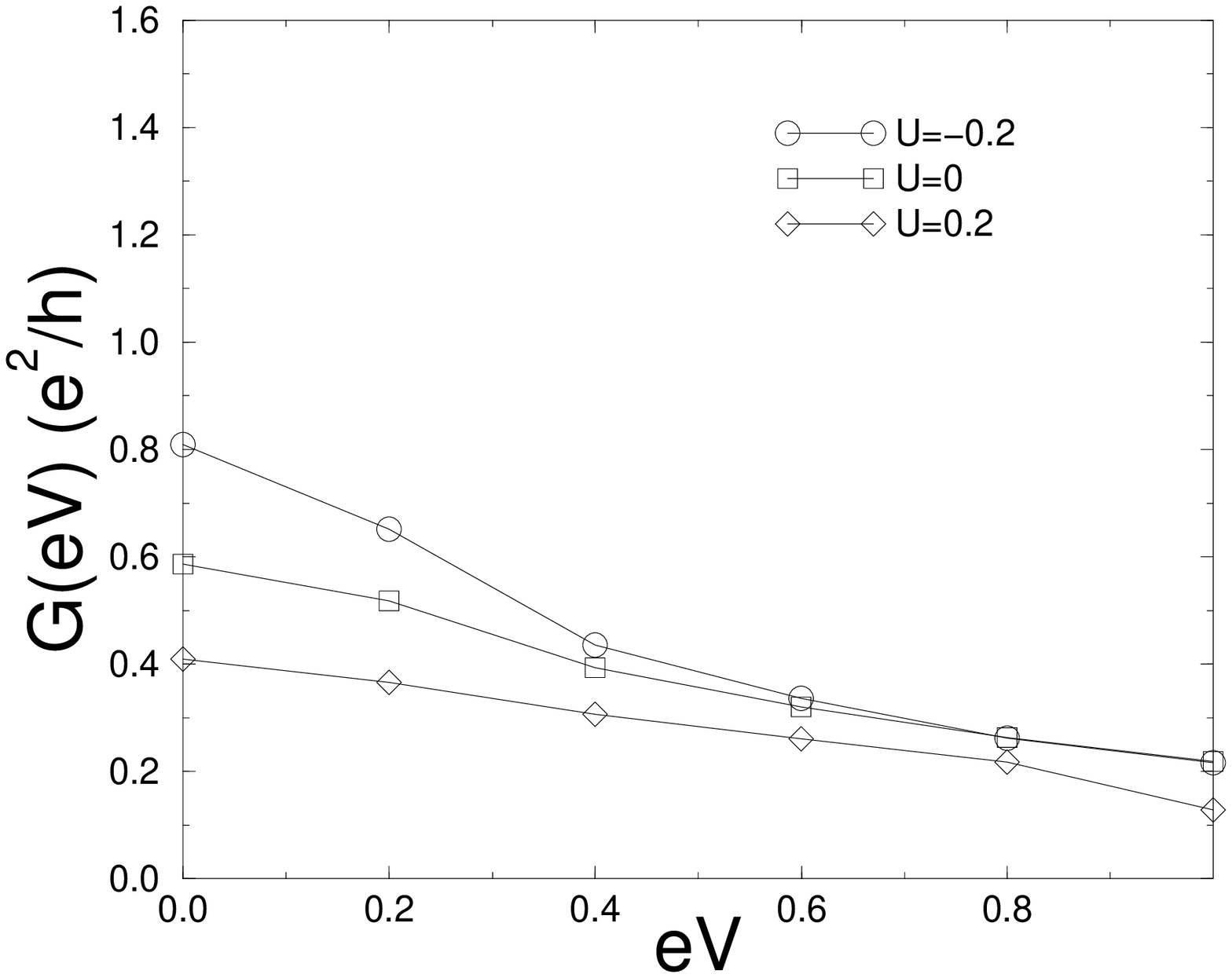}
\end{center}
\caption{
Differential conductance evaluated by means of the approximate formula 
({\protect \ref{differential_conductance_WML}}) with our spectral-function data
shown in Fig. {\protect \ref{rho_Vt3_Ut001_10}}-{\protect \ref{rho_Vt3_Umt2_10}}.
This is to be compared with the first-principle result shown in
Fig. {\protect \ref{ogon_Vt3}}.
}
\label{caroli_Vt3}
\end{figure}

Finally, we show the result of the differential conductance
based on the approximative formula (\ref{differential_conductance_WML}).
The application of the formula is now possible, because 
we have nonequilibrium-Green-function data.
The result is plotted in Fig. \ref{caroli_Vt3},
Although this approximative formula reproduces 
the zero-bias anomaly, the dependence on $U$ differs significantly 
from 
that of our first-principle result.
For instance, in Fig. \ref{caroli_Vt3}, the conductance
is {\em increased} by $U<0$, whereas it should be suppressed 
according to our simulation.
(Our conclusion may also be supported by a recent report,
where, however, the authors studied the tunneling amplitude
from one edge to the bulk of the Hubbard chain \cite{Bedurftig99}.)
It is suspected that the approximate formula
does not fully include the many-body correlation effect.
One cannot expect, as a matter of fact, that
the nonequilibrium transport coefficient is
given by one-particle Green function alone,
because the Kubo theory 
tells that the {\em non}-linear-response function should be expressed 
in terms of 
{\em many}-point Green functions.

\section{Summary}
\label{section_summary}
We have investigated numerically the nonequilibrium electron transport through
the Kondo impurity.
In this problem, nonequilibrium-driving force and biquadratic
correlation coexist, and both are playing a crucial role.
For the first time, we succeeded in simulating the situation
without resorting to any perturbative treatments.
Moreover, because we utilized the mapping to the resonance
level model, 
we could clarify the effect of the biquadratic correlation
as a gradual deviation from the Toulouse point.
We calculated the differential conductance with the
formula (\ref{differential_conductance})
and the nonequilibrium Green function with eq. (\ref{Green_function}). 
Our results tell that the effect of the biquadratic interaction lies
out of the scope of the one-particle description:
The attractive density-density coupling, which
corresponds to a reduced Kondo coupling from the Toulouse point,
suppresses the zero-bias
conductance below the Landauer value.
The width of the zero-bias anomaly, on the other hand, is not changed
very much.
The above is to be contrasted with the field-theoretical
lowest-order-perturbation theory 
with respect to the Toulouse point \cite{Majumdar98}.
The repulsive coupling, on the contrary,
does not influence the conductance.
The above features are supported by
our
spectral function data based on eq. (\ref{DOS}).

We stress that as we have demonstrated, the `steady nonequilibrium'
transport can be simulated without carrying out any time-evolution
simulations.
The formalism and the implementation to the computer algorithm,
which are
demonstrated here,
are readily applicable to other wide class of steady
nonequilibrium transport with many-body interaction.
As is clarified here, in such situation as well,
the many-body correlation would cause new unexpected behavior.
This would be remained in future.

\section*{Acknowledgments}
The author is grateful to Prof. W. Apel and 
Prof. H.-U. Everts for helpful discussions.
Hospitality at Institut f\"ur Theoretische Physik, Universit\"at
Hannover, is gratefully acknowledged.

% \begin{figure}[htbp]
% \begin{center}\leavevmode
% \epsfxsize=17cm
% \epsfbox{ogon_Vt4.eps}
% \end{center}
% \caption{
% Differential conductance (\ref{differential_conductance})
% is plotted for $V_{\rm h}=0.4$ and various $U$.
% }
% \label{ogon_Vt4}
% \end{figure}

\begin{thebibliography}{99}
\bibitem{Goldhaber-Gordon98}
D. Goldhaber-Gordon, H. Shtrikman, D. Mahalu,
D. Abusch-Magder, U. Meirav and M. A. Kastner:
Nature {\bf 391} (1998) 156.
\bibitem{Cronenwett98}
S. M. Cronenwett, T. H. Oosterkamp and L. P. Kouwenhoven:
Science {\bf 281} (1998) 540.
\bibitem{Schmid98}
J. Schmid, J. Weis, K. Eberl and K. v. Klitzing:
Physica B {\bf 256-258} (1998) 182.
\bibitem{Wyatt64}A. F. G. Wyatt: Phys. Rev. Lett. {\bf 13} (1964) 401.
\bibitem{Logan64}R. A. Logan and J. M. Rowell: Phys. Rev. Lett. {\bf 13} (1964) 404.
\bibitem{Gregory92}S. Gregory: Phys. Rev. Lett. {\bf 68} (1992) 2070.
\bibitem{Ralph94}D. C. Ralph and R. A. Buhrman:
Phys. Rev. Lett. {\bf 72} (1994) 3401.
\bibitem{Appelbaum66}J. Appelbaum: Phys. Rev. Lett. {\bf 17} (1966) 91.
\bibitem{Anderson66}P. W. Anderson: Phys. Rev. Lett. {\bf 17} (1966) 95.
\bibitem{Glazman88}L. I. Glazman and M. \'E. Raikh:
JETP Lett. {\bf 47} (1988) 452.
\bibitem{Ng88}T. K. Ng and P. A. Lee: Phys. Rev. Lett. {\bf 61} (1988) 1768.
\bibitem{Kawabata91}A. Kawabata: J. Phys. Soc. Japan {\bf 60} (1991) 3222.
\bibitem{Kawabata98}A. Kawabata: J. Phys. Soc. Japan {\bf 67} (1998) 2430
\bibitem{Meir91}Y. Meir, N. S. Wingreen and P. A. Lee:
Phys. Rev. Lett. {\bf 66} (1991) 3048.
\bibitem{Meir92}Y. Meir and N. S. Wingreen: Phys. Rev. Lett. {\bf 68} (1992) 2512.
\bibitem{Meir93}Y. Meir, and N. S. Wingreen and P. A. Lee:
Phys. Rev. Lett. {\bf 70} (1993) 2601.
\bibitem{Ng93}T. K. Ng: Phys. Rev. Lett. {\bf 70} (1993) 3635.
\bibitem{Wingreen94}N. S. Wingreen and Y. Meir: Phys. Rev. B {\bf 49} (1994) 11040.
\bibitem{Hershfield91}S. Hershfield, J. H. Davies and J. W. Wilkins:
Phys. Rev. Lett. {\bf 67} (1991) 3720. 
\bibitem{Hershfield92}S. Hershfield, J. H. Davies and J. W. Wilkins:
Phys. Rev. B {\bf 46} (1992) 7046.
\bibitem{Schiller95}A. Schiller and S. Hershfield: Phys. Rev. B {\bf 51} (1995) 12896.
\bibitem{Majumdar98}K. Majumdar, A. Schiller and S. Hershfield.
Phys. Rev. B {\bf 57} (1998) 2991.
\bibitem{Schiller98}A. Schiller and S. Hershfield: Phys. Rev. B {\bf 58} (1998) 14978.
\bibitem{Inoshita93}T. Inoshita, A. Shimizu, Y. Kuramoto and H. Sakaki:
Phys. Rev. B {\bf 48} (1993) 14725.
\bibitem{Izumida98}W. Izumida, O. Sakai and Y. Shimizu: J. Phys. Soc. Japan
{\bf 67} (1998) 2444.
\bibitem{Oguri95}A. Oguri, H. Ishii and T. Saso: Phys. Rev. B {\bf 51} (1995) 4715.
\bibitem{Oguri97}A. Oguri: Phys. Rev. B {\bf 56} (1997) 13422; (E) {\bf 58} (1998) 1690.
\bibitem{Konig96}J. K\"onig, H. Schoeller and G. Sch\"on:
Phys. Rev. Lett. {\bf 76} (1996) 1715.
\bibitem{Caroli71}C. Caroli, R. Combescot, P. Nozieres and D. Saint-James:
J. Phys. C: Solid St. Phys. {\bf 4} (1971) 916.
\bibitem{Keldysh65}L. V. Keldysh: Sov. Phys.-JETP {\bf 20} (1965) 1018. 
\bibitem{Toulouse70}G. Toulouse: Phys. Rev. B {\bf 2} (1970) 270.
\bibitem{Emery92}V. J. Emery and S. A. Kivelson: Phys. Rev. B {\bf 46} (1992) 10812.
%\bibitem{Landauer87}R. Landauer: Z. Phys. B {\bf 68} (1987) 217.
\bibitem{Landauer57}R. Landauer: IBM J. Res. Dev. {\bf 1} (1957) 223.
\bibitem{Imry99}Y. Imry and R. Landauer: Rev. Mod. Phys. {\bf 71} (1999) S306.
% \bibitem{Schotte70}K. D. Schotte: Z. Phys. {\bf 230} (1970) 99.
\bibitem{Schlottmann82}P. Schlottmann: Phys. Rev. B {\bf 25} (1982) 4815.
\bibitem{Guinea85}F. Guinea, V. Hakim and A. Muramatsu:
Phys. Rev. B {\bf 32} (1985) 4410.
\bibitem{Gell-Mann51}M. Gell-Mann and F. Low: Phys. Rev. {\bf 84} (1951) 350.
\bibitem{Hershfield93}S. Hershfield: Phys. Rev. Lett. {\bf 70} (1993) 2134.
\bibitem{Gagliano87}E. R. Gagliano, C. A. Balseiro: Phys. Rev. Lett. {\bf 59} (1987) 2999.
\bibitem{Bedurftig99}G. Bed\"urftig and H. Frahm: preprint.




\end{thebibliography}
\end{document}